\documentclass[journal]{IEEEtran}

\usepackage{graphicx}  

\usepackage{amsmath}   
\usepackage{amssymb}

\newcommand{\bc}{\begin{center}}
\newcommand{\ec}{\end{center}}
\newcommand{\bt}{\begin{tabular}}
\newcommand{\et}{\end{tabular}} 
\newcommand{\bea}{\begin{eqnarray}}
\newcommand{\eea}{\end{eqnarray}}
\newcommand{\bean}{\begin{eqnarray*}}
\newcommand{\eean}{\end{eqnarray*}}

\newcommand{\ba}{\begin{array}}
\newcommand{\ea}{\end{array}}

\def\be{\begin{eqnarray}}
\def\ee{\end{eqnarray}}
\def\ben{\begin{eqnarray*}}
\def\een{\end{eqnarray*}}

\newcommand{\ra} {\rightarrow}

\newcommand{\leqb}{\mbox{$ \;\stackrel{(b)}{\leq}\; $}}

\newcommand{\leqe}{\mbox{$ \;\stackrel{(e)}{\leq}\; $}}

\newcommand{\geqb}{\mbox{$ \;\stackrel{(b)}{\geq}\; $}}

\newcommand{\geqd}{\mbox{$ \;\stackrel{(d)}{\geq}\; $}}

\newcommand{\eqa}{\mbox{$ \;\stackrel{(a)}{=}\; $}}

\newcommand{\eqc}{\mbox{$ \;\stackrel{(c)}{=}\; $}}

\newcommand{\nth}{\frac{1}{n}}

\newcommand{\RL}{{\mathbb R}}

\newcommand{\calF}{\mbox{${\cal F}$}}

\newcommand{\calP}{\mbox{${\cal P}$}}

\newcommand{\calX}{\mbox{${\cal X}$}}

\newcommand{\Rpl}{{\mathbb R}_{+}}
\newcommand{\Zpl}{\mathbb{Z}_{+}}

\newcommand{\Sumn}{\sum_{i=1}^{n}}

\newcommand{\Ent}{\mbox{Ent}}

\def\elabel#1{\label{e:#1}}

\def\sq{$\Box$}

\def\qed{\ifmmode\sq\else{\unskip\nobreak\hfil
\penalty50\hskip1em\null\nobreak\hfil\sq
\parfillskip=0pt\finalhyphendemerits=0\endgraf}\fi\par\medbreak}

\newsavebox{\junk}
\savebox{\junk}[1.6mm]{\hbox{$|\!|\!|$}}

\newcommand{\one}{\hbox{\rm\large\textbf{1}}}

\def\til={{\widetilde =}}

\def\half{{\mathchoice{\textstyle \frac{1}{2}}%
{\frac{1}{2}}%
{\hbox{\tiny $\frac{1}{2}$}}%
{\hbox{\tiny $\frac{1}{2}$}} }}

 \def\eq#1/{(\ref{#1})}

\def\eq#1/{(\ref{e:#1})}

\newcommand{\beqn}[1]{\notes{#1}%
\begin{eqnarray} \elabel{#1}}

\newcommand{\eeqn}{\end{eqnarray} }

\newcommand{\beq}[1]{\notes{#1}%
\begin{equation}\elabel{#1}}

\newcommand{\eeq}{\end{equation}} 

\def\bdes{\begin{description}}
\def\edes{\end{description}}

\def\notes#1{}

\newcommand{\calS}{\mathcal{C}}
\newcommand{\calN}{\mathcal{N}}
\newcommand{\setS}{s}
\newcommand{\setT}{t}
\newcommand{\setU}{u}
\newcommand{\collS}{\calS}
\newcommand{\as}{\alpha(\setS)}
\newcommand{\bs}{\beta(\setS)}
\newcommand{\gs}{\gamma(\setS)}
\newcommand{\Xs}{X_{\setS}}
\newcommand{\Xsc}{X_{\setS^{c}}}
\newcommand{\Xt}{X_{\setT}}
\newcommand{\ws}{w_{\setS}}

\newcommand{\rth}{\frac{1}{r}}
\newcommand{\gap}{\text{Gap}}
\newcommand{\sumS}{\sum_{\setS\in\collS}}

\renewcommand{\calP}{\mathbb{P}}
\newcommand{\calQ}{\mathbb{Q}}
\newcommand{\e}{{\tt e}}

\def\I{{\cal I}}

\def\R{{\bf R}}

\newcommand{\nullset}{\phi}
\newcommand{\Hom}{\text{Hom}}

\begin{document}
%
\title{Information Inequalities for Joint Distributions, with Interpretations and Applications}
\author{Mokshay~Madiman,~\IEEEmembership{Member,~IEEE,}
        and~Prasad~Tetali,~\IEEEmembership{Member,~IEEE}
\thanks{Material in this paper was presented 
at the Information Theory and Applications Workshop, San Diego, CA, January 2007,
and at the IEEE Symposium on Information Theory, Nice, France, June 2007.}%
\thanks{Mokshay Madiman is with 
the Department of Statistics, Yale University,
24 Hillhouse Avenue, New Haven, CT 06511, USA.
Email: {\tt mokshay.madiman@yale.edu}}
\thanks{Prasad Tetali is with the School
of Mathematics and College of Computing, Georgia Institute of
Technology, Atlanta, GA 30332, USA. Email:
\texttt{tetali@math.gatech.edu}. Supported in part by NSF grants
DMS-0401239 and DMS-0701043.}%
}
%
%
%
\markboth{Submitted to IEEE Transactions on Information Theory,~2007}{Madiman and Tetali \MakeLowercase
: Information Inequalities}
%


\maketitle

\begin{abstract}
Upper and lower bounds are obtained for the joint entropy of a collection of random variables
in terms of an arbitrary collection of subset joint entropies. These inequalities generalize 
Shannon's chain rule for entropy as well as inequalities of Han, Fujishige and Shearer.
A duality between the upper and lower bounds for joint entropy is developed.
All of these results are shown to be special cases of general, new results for submodular
functions-- thus, the inequalities presented constitute a richly structured class of 
Shannon-type inequalities.
The new inequalities are applied to obtain new results in combinatorics, 
such as bounds on the number of independent 
sets in an arbitrary graph and the number of zero-error source-channel codes,
as well as new determinantal inequalities in matrix theory. A new inequality for
relative entropies is also developed, along with interpretations in terms of  
hypothesis testing. Finally, revealing connections of the results to literature in 
economics, computer science, and physics are explored. 
\end{abstract}


%

\section{Introduction}
\label{sec:intro}

\PARstart{L}{et}  
 $X_1, X_2, \ldots,X_{n}$ be a collection of
random variables. There are the familiar two canonical cases:
(a) the random variables are real-valued and possess a
probability density function, in which case
$h$ represents the differential entropy, 
or (b) they are discrete, in which case
$H$ represents the discrete entropy. 
More generally, if the joint distribution has
a density $f$ with respect to some reference product measure, 
the joint entropy may be defined by 
$-E[\log f(X_1, X_2, \ldots,X_{n})]$; with this definition,
$H$ corresponds to counting measure and $h$ to Lebesgue measure.
The only assumption we will implicitly make throughout is
that the joint entropy 
is finite, i.e., neither $-\infty$ nor $+\infty$.

We wish to discuss the relationship between the joint entropies
of various subsets of the random variables $X_1, X_2, \ldots,X_{n}$.
Thus we are motivated to consider an arbitrary collection $\collS$ 
of subsets of $\{1,2,\ldots, n\}$.
The following conventions are useful:
\begin{itemize}
\item
$[n]$ is the index set $\{1,2,\ldots,n\}$. We equip this set with its
natural (increasing) order, so that $1<2<\ldots<n$.
(Any other total order would do equally well, and indeed we use this
flexibility later, but it is convenient to fix a default order.)
\item
For any set $\setS\subset [n]$,
$\Xs$ stands for the collection of random
variables $(X_{i}:i\in \setS)$, with the indices taken
in their increasing order.
\item
For any index $i$ in $[n]$, define the {\it degree}
of $i$ in $\collS$ as $r(i)=|\{\setT\in\collS: i\in \setT\}|$.
Let $r_{-}(\setS)=\min_{i\in\setS} r(i)$ denote the
{\it minimal degree} in $\setS$, and
$r_{+}(\setS)=\max_{i\in\setS} r(i)$
denote the {\it maximal degree} in $\setS$. 
\end{itemize}

First we present a weak form of our main inequality.

\vspace{.13in}\noindent{\bf Proposition I:}[{\sc Weak degree form}]
Let $X_1, \ldots, X_n$ be arbitrary random variables jointly
distributed on some discrete sets.
For any collection $\collS$ such that each index $i$
has non-zero degree,
\be\label{prop1}
\sumS \frac{H(\Xs|\Xsc)}{r_{+}(\setS)} 
\leq H(X_{[n]}) \leq \sumS \frac{H(\Xs)}{r_{-}(\setS)}  ,
\ee
where $r_{+}(\setS)$ 
and  $r_{-}(\setS)$ 
are the maximal and minimal degrees in $\setS$. If $\collS$
satisfies $r_{-}(\setS)=r_{+}(\setS)$ for each $\setS$ in $\collS$,
then \eqref{prop1} also holds for $h$ in the setting of continuous random variables.
\vspace{.13in}

Proposition I unifies a large number of inequalities in the literature.
Indeed,
\begin{enumerate}
\item
\noindent Applying to the class $\collS_{1}$ of singletons, 
\be\label{weak:s1}
\Sumn H(X_{i}|X_{[n]\setminus i}) \leq H(X_{[n]}) \leq \Sumn H(X_{i}) .
\ee
The upper bound represents the subadditivity of entropy noticed by Shannon.
The lower bound may be interpreted as the fact that the erasure entropy
of a collection of random variables is not greater than their entropy; see
Section VI for further comments.
\item
\noindent Applying to the class $\collS_{n-1}$ of all sets of $n-1$ elements,
\be\begin{split}\label{han:proto}
\frac{1}{n-1} \Sumn H(X_{[n]\setminus i}&|X_{i}) \leq H(X_{[n]})\\
&\leq \frac{1}{n-1} \Sumn H(X_{[n]\setminus i}) .
\end{split}\ee
This is Han's inequality \cite{Han78, CT91:book}, in its prototypical form.
\item
\noindent Let $r_{+}=\min_{i\in[n]} r(i)$ and 
$r_{-}=\max_{i\in[n]} r(i)$ be the minimal and maximal
degrees with respect to $\collS$.
Using $r_{-}\leq r_{-}(\setS)$ and $r_{+}\leq r_{+}(\setS)$, we have
\ben
\frac{1}{r_{+}} \sumS H(X_{\setS}|\Xsc) 
\leq H(X_{[n]}) \leq 
\frac{1}{r_{-}} \sumS H(X_{\setS}) .
\een
The upper bound is Shearer's lemma \cite{CGFS86}, known
in the combinatorics literature \cite{Rad03}. The lower bound 
is new.
\end{enumerate}

The paper is organized as follows. First, in 
Section~\ref{sec:hyp}, the notions of fractional coverings and 
packings using hypergraphs, which provide a useful language for 
the information inequalities we present, are developed.
In Section~\ref{sec:submod}, we present the main technical result
of this paper, which is a new inequality for submodular set functions.
Section~\ref{sec:main} presents the main entropy inequality of this paper, which
strengthens Proposition I, and gives a very simple proof as a corollary of the
general result for submodular functions. This entropy inequality is developed in
two forms, which we call the strong fractional form and the 
strong degree form; Proposition I may then be thought of
as the weak degree form.
A different manifestation of the upper bound in this weak degree form
of the inequality was recently proved (in a more involved manner) by 
Friedgut \cite{Fri04}; the relationship with his result is also
further discussed in Section~\ref{sec:main} using 
the preliminary concepts developed in Section~\ref{sec:hyp}.

While independent sets in graphs have always been of
combinatorial and graph-theoretical interest, counting independent
sets in bipartite graphs received renewed attention due to Kahn's
entropy approach \cite{Kah01a} to Dedekind's problem. Dedekind's
problem involves counting the number of antichains in the Boolean lattice, 
or equivalently, counting the number of Boolean functions on $n$ variables
that can be constructed using only AND and OR (and no NOT) gates.
To handle this problem by induction on the number of levels in the
lattice,  Kahn first derived a tight bound on the (logarithm of
the) number of independent sets in a regular bipartite graph. 
In Section~\ref{sec:count}, we build on Kahn's work to
obtain a bound on number of independent sets in an arbitrary graph.
We also generalize this to counting graph homomorphisms,
with applications to graph coloring and zero-error source-channel codes.

The applications of entropy inequalities to counting typically
involves discrete random variables, but the inequalities also
have applications when applied to continuous random variables. 
In Section~\ref{sec:det}, we develop such an application by
proving a new family of determinantal inequalities that provide
generalizations of the classical determinantal inequalities of 
Hadamard, Szasz and Fischer.

Having presented two applications of our main inequalities,
we move on to studying the structure of the inequalities more
closely. In Section~\ref{sec:dual}, we present a duality between our upper
and lower bounds that generalizes a theorem of Fujishige \cite{Fuj78}. In particular, we
show that the collection of upper bounds on $H(X_{[n]})$ for all
collections $\collS$ is equivalent to the collection of lower bounds.
There we also discuss interpretations of the inequality relating to sensor networks and
erasure entropy, and generalize the monotonicity property for special collections
of subsets discovered by Han \cite{Han78}.

Section~\ref{sec:epi} presents some new entropy power
inequalities for joint distributions, and points out an intriguing analogy 
between them and the recent subset sum entropy power inequalities
of Madiman and Barron \cite{MB07}.
In Section~\ref{sec:D}, we prove inequalities for relative entropy between
joint distributions. Interpretations of the relative entropy inequality through 
hypothesis testing and concentration of measure are also given there. 

In Section~\ref{sec:hist}, we note that weaker versions of our main inequality 
for submodular functions follow from results developed in various communities 
(economics, computer science, physics); this history and the consequent connections 
do not seem to be well known or much tapped in information theory. Finally in
Section~\ref{sec:disc}, we conclude with some final remarks and 
brief discussion of other applications, including to multiuser information theory.

\section{On Hypergraphs and Related Concepts}
\label{sec:hyp}

It is appropriate here to recall some terminology from discrete
mathematics. A collection $\collS$ of subsets of $[n]$ is called
a {\it hypergraph}, and each set $\setS$ in $\collS$ is called
a {\it hyperedge}. When each hyperedge has cardinality 2,
then $\collS$ can be thought of as the set of edges
of an undirected graph on $n$ labelled vertices. Thus all the statements made above
can be translated into the language of hypergraphs. 
In the rest of this paper, we interchangeably use ``hypergraph'' and ``collection''
for $\collS$,  ``hyperedge'' and ``set'' for $\setS$ in $\collS$,
and ``vertex'' and ``index'' for $i$ in $[n]$.

We have the following standard definitions.

\vspace{.13in}\noindent{\bf Definition I:}
The collection $\collS$ is said to be {\em $r$-regular} if
each index $i$ in $[n]$ has the same degree $r$, i.e.,
if each vertex $i$ appears in exactly $r$ hyperedges of $\collS$.
\par\vspace{.13in}

The following definitions extend the familiar notion of 
packings, coverings and partitions of sets by allowing 
fractional counts. The history of these notions is unclear
to us, but some references can be found in the book by
Scheinerman and Ullman \cite{SU97:book}.

\vspace{.13in}\noindent{\bf Definition II:}
Given a collection $\collS$ of subsets of $[n]$, a function $\alpha:\collS \to \R^+$,
is called a {\em fractional covering}, if for each $i\in [n]$, we have
$\sum_{\setS\in \collS:i\in \setS} \as \ge 1$.

Given $\collS$, a function $\beta:\collS \to \R^+$ 
is a {\em fractional packing}, if  for each $i\in [n]$, we have
$\sum_{\setS\in \collS:i\in S} \bs \le 1$. 

If $\gamma:\collS\to\R^{+}$ is both a fractional covering and a fractional
packing, we call $\gamma$ a {\em fractional partition}.
\par\vspace{.13in}

Note that the standard definition of a fractional packing
of $[n]$ using $\collS$ (as in \cite{SU97:book}), would assign weights $\beta_i$ to the elements,
(rather than sets) $i\in [n]$, and require that, for each $\setS\in\collS$,
we have $\sum_{i \in\setS} \beta_i \le 1$. Our terminology can be justified, if
one considers the ``dual hypergraph," obtained by interchanging
the role of elements and sets -- consider the 0-1 incidence matrix (with
rows indexed by the elements and columns by the sets) of the set
system, and simply switch the roles of the elements and the sets.

The following simple lemmas are useful.

\vspace{.13in}\noindent{\bf Lemma I:}[{\sc Fractional Additivity}]
Let $\{a_{i}:i\in [n]\}$ be an arbitrary collection of real numbers. 
For any $\setS\subset [n]$, define $a_{\setS}=\sum_{j\in\setS} a_{j}$.
For any fractional partition $\gamma$ using any hypergraph $\collS$, 
$a_{[n]} = \sumS \gamma(\setS) a_{\setS}$. 
Furthermore, if each $a_{i}\geq 0$, then
\be\label{f1}
\sumS \bs a_{\setS} \leq a_{[n]} \leq
\sumS \as a_{\setS} 
\ee
for any fractional packing $\beta$ and any fractional covering $\alpha$
using $\collS$.
\par\vspace{.13in}

\begin{proof}
Interchanging sums implies
\ben\begin{split}
\sumS \as \sum_{i\in \setS} a_{i}
= \sum_{i\in [n]} a_{i} \sumS \as {\bf 1}_{\{i\in\setS\}}
\geq \sum_{i\in [n]} a_{i} ,
\end{split}\een
using the definition of a fractional covering. 
The other statements are similarly obvious. 
\end{proof}

We introduce the notion of quasiregular hypergraphs.

\vspace{.13in}\noindent{\bf Definition III:}
The hypergraph $\collS$ is {\em quasiregular} if the degree function
$r:[n]\ra\Zpl$ defined by $r(i)=|\{\setS\in\collS:\setS\ni i\}|$ is
constant on $\setS$, for each $\setS$ in $\collS$.
\par\vspace{.13in}

\noindent{\em Example:}
One can construct simple examples of quasiregular hypergraphs 
using what are called bi-regular graphs in the graph theory literature.
Consider a bipartite graph on vertex sets $V_1$ and $V_2$
(i.e., all edges go between $V_1$ and $V_2$), such that
every vertex in $V_1$ has degree $r_1$ and every vertex in $V_2$ has degree
$r_2$. Such a graph always exists if $|V_{1}| r_{1}=|V_{2}| r_{2}$.
Now consider the hypergraph on $V_1\cup V_2$ with hyperedges being
the neighborhoods of vertices in the bipartite graph. This hypergraph
is quasiregular (with degrees being $r_1$ and $r_2$), and it is not regular if
$r_1$ is different from $r_2$.
\par\vspace{.13in}

There is a sense in which all quasiregular hypergraphs are
similar to the example above; specifically, 
any quasiregular hypergraph has a canonical decomposition as
a disjoint union of regular subhypergraphs.

\vspace{.13in}\noindent{\bf Lemma II:}
Suppose the hypergraph $\collS$ on the vertex set $[n]$ is quasiregular.
Then one can partition $[n]$ into disjoint subsets $\{V_{m}\}$, and $\collS$
into disjoint subhypergraphs $\{\collS_{m}\}$ such that each $\collS_{m}$
is a regular hypergraph on vertex set $V_{m}$.
\par\vspace{.13in}

\begin{proof}
Consider the equivalence relation on $[n]$ induced by the degree, i.e.,
$i$ and $j$ are related if $r(i)=r(j)$. This relation decomposes $[n]$
into disjoint equivalence classes $\{V_{m}\}$. 
Since $\collS$ is quasiregular, all indices in $\setS$ have the same degree 
for each set $\setS\in\collS$, and hence each $\setS\in\collS$
is a subset of exactly one equivalence class $V_{m}$. Q.E.D.
\end{proof}
\par\vspace{.13in}

The notion of quasiregularity is related to what we believe is an
important and natural fractional covering/packing pair.
As long as there is at least one set $\setS$ in the hypergraph $\collS$ 
that contains $i$, we have
\ben\begin{split}
\sum_{\setS\in\collS, \setS\ni i} \frac{1}{r_{-}(\setS)} 
= \sumS \frac{{\bf 1}_{\{i\in\setS\}}}{r_{-}(\setS)} 
\geq  \sumS \frac{{\bf 1}_{\{i\in\setS\}}}{r(i)}
= 1 ,
\end{split}\een
so that  $\as=\frac{1}{r_{-}(\setS)}$
provide a fractional covering. Similarly, the 
the numbers $\bs=\frac{1}{r_{+}(\setS)}$
provide a fractional packing.

\par\vspace{.13in}
\noindent{\bf Definition IV:}
Let $\collS$ be any hypergraph on $[n]$ such that
every index appears in at least one hyperedge. 
The fractional covering given by $\as=\frac{1}{r_{-}(\setS)}$
is called the {\it degree covering}, and the fractional packing
given by $\bs=\frac{1}{r_{+}(\setS)}$
is called the {\it degree packing}.
\par\vspace{.13in}

The following lemma is a trivial consequence of the definitions. 

\vspace{.13in}\noindent{\bf Lemma III:}
If $\collS$ is quasiregular, the degree packing and degree covering
coincide and provide a fractional partition of $[n]$ using $\collS$.
In particular, $a_{[n]} = \sumS a_{\setS}/r_{-}(\setS)$.


One may define the weight of a fractional partition as follows.

\par\vspace{.13in}
\noindent{\bf Definition V:}
Let $\gamma$ be a fractional partition (or a fractional covering or packing).
Then the weight of $\gamma$ is 
$w(\gamma)=\sumS  \gs$.
\par\vspace{.13in}

There are natural optimization problems associated with the weight function.
The problem of minimizing the weight 
of $\alpha$ over all fractional coverings $\alpha$
is the called the  {\it optimal fractional covering} problem, 
and that of maximizing the weight 
of $\beta$ over all fractional packings $\beta$
is the called the  {\it optimal fractional packing} problem. 
These are linear programming relaxations of the integer
programs associated with optimal covering and optimal packing,
which are of course important in many applications. Much work
has been done on these problems, including studies of the integrality gap
(see, e.g., \cite{SU97:book}).


One may also define a notion of duality for fractional partitions.

\par\vspace{.13in}
\noindent{\bf Definition VI:}
For any hypergraph $\collS$, define the complimentary hypergraph as
$\bar{\collS}=\{\setS^{c}:\setS\in\collS\}$. If $\alpha$ is a fractional covering (or packing)
using $\collS$, the dual  fractional packing (respectively, covering) using $\bar{\collS}$ 
is defined by
\ben
\bar{\alpha}(\setS^{c}) =\frac{\as}{w(\alpha) - 1} .
\een
\par\vspace{.13in}

To see that this definition makes sense (say for the case of a fractional covering $\alpha$), 
note that for each $i\in [n]$,
\ben\begin{split}
\sum_{\setS^{c}\in \bar{\collS}, \setS^{c}\ni i} \bar{\alpha}(\setS^{c}) 
&= \sum_{\setS\in\collS, i\notin \setS}  \frac{\as}{w(\alpha) - 1} \\
&= \frac{\sumS \as - \sum_{\setS\in\collS, i\in \setS} \as}{w(\alpha) - 1} \\
&\leq \frac{w(\alpha) - 1}{w(\alpha) - 1} =1.
\end{split}\een
\par\vspace{.05in}

\section{A new inequality for submodular functions}
\label{sec:submod}

The following definitions are necessary in order to state the
main technical result of this paper. 

\vspace{.13in}\noindent{\bf Definition VII:}
The set function $f:2^{[n]}\ra\RL$ is {\em submodular} if 
\ben
f(\setS) + f(\setT) \ge f(\setS\cup\setT) + f(\setS\cap\setT)
\een
for every $\setS, \setT \subset [n]$. If $-f$ is submodular,
we say that $f$ is {\em supermodular}.

\vspace{.13in}\noindent{\bf Definition VIII:}
For any disjoint subsets $\setS$ and $\setT$ of $[n]$, define
$f(\setS|\setT)=f(\setS\cup\setT)-f(\setT)$. For a fixed subset $\setT\subsetneq [n]$,
the function $f_{\setT}:2^{[n]\setminus\setT}\ra\RL$ defined
by $f_{\setT}(\setS)=f(\setS|\setT)$ is called {\em $f$ conditional on $\setT$}.
\par\vspace{.13in}

For any $\setS\subset [n]$, denote by $<\setS$ the set of indices less than
every index in $\setS$. Similarly, $>\setS$ is the set of indices greater than
every index in $\setS$.
Also, the index $i$ is identified with the set $\{i\}$;
thus, for instance, $<i$ is well-defined. We also
write $[i:i+k]$ for $\{i,i+1,\ldots,i+k-1,i+k\}$. Note that
$[n]=[1:n]$.

\vspace{.13in}\noindent{\bf Lemma IV:} 
Let $f:2^{[n]}\ra\RL$ be any submodular function with $f(\nullset)=0$.
\begin{enumerate}
\item If $\setS, \setT, \setU$ are disjoint sets, 
\be\label{condred}
f(s|t,u) \leq f(s|t).
\ee
\item The following ``chain rule'' expression holds for $f([n])$:
\ben
f([n])=\sum_{i\in[n]} f(i|<i) .
\een
\end{enumerate}
\par\vspace{.13in}
\begin{proof}
First note that if $\setS, \setT, \setU$ are disjoint sets, then submodularity implies
\ben
f(\setS\cup\setT\cup\setU)+f(\setT) \leq f(\setS\cup\setT)+f(\setT\cup\setU) ,
\een
which is equivalent to $f(s|t,u) \leq f(s|t)$.

The ``chain rule'' expression for $f([n])$ is obtained by induction. Note
that $f([2])=f(1)+f(2|1)= f(1|\nullset)+f(2|1)$ since $f(\nullset)=0$. Now assume
the chain rule holds for $[n]$, and observe that
\ben\begin{split}
f([n+1])=f([n])+f(n+1|[n])
= \sum_{i\in[n+1]} f(i|<i) ,
\end{split}\een
where we used the induction hypothesis for the second equality.
\end{proof}

\vspace{.13in}\noindent{\bf Theorem I:} 
Let $f:2^{[n]}\ra\RL$ be any submodular function with $f(\nullset)=0$.
Let $\gamma$ be any fractional partition with respect to any collection $\collS$ of subsets of $[n]$.
Then
\ben\label{mainH}
\sumS \gs f (\setS | \setS^{c}\setminus >\setS) 
\leq f([n]) \leq  
\sumS \gs f (\setS | <\setS)\, .
\een
\par\vspace{.13in}

\begin{proof}
The chain rule (actually a slightly extended version of it with additional conditioning in all terms
that can be proved in exactly the same way) implies
\be\label{xs-chain}
f(\setS | <\setS)=\sum_{j\in\setS} f(j| <j\cap\setS , <\setS ) .
\ee
Thus
\ben\begin{split}
\sumS  \as  f(\setS | <\setS)
&\eqa \sumS \as \sum_{j\in\setS} f(j| <j\cap\setS,  <\setS ) \\
&\geqb \sumS \as \sum_{j\in\setS} f(j| <j) \\
&\eqc \sum_{j\in [n]} f(j| <j) \sumS \as {\bf 1}_{\{j\in\setS\}} \\
&\geqd \sum_{j\in [n]} f(j| <j) \\
&\eqa f(X_{[n]}) ,
\end{split}\een
where (a) follows by the chain rule \eqref{xs-chain}, (b) follows from \eqref{condred},
(c) follows by interchanging sums, and (d) follows by the definition of a fractional covering.

The lower bound may be proved in a similar fashion by a chain of inequalities. 
Indeed,
\ben\begin{split}
&\sumS \bs f (\setS | \setS^{c}\setminus >\setS) \\
&\eqa \sumS \bs
\sum_{j\in\setS} f(j| <j\cap\setS,  \setS^{c}\setminus >\setS )   \\
&\leqb \sumS \bs \sum_{j\in\setS} f(j| <j )   \\
&\eqc \sum_{j\in [n]} f(j| <j) \sumS {\bf 1}_{\{j\in\setS\}} \bs \\
&\leqe \sum_{j\in [n]} f(j| <j) \\
&\eqa f([n]) ,
\end{split}\een
where (a), (b), (c) follow as above, and (e) follows by the definition of a fractional partition.
\end{proof}

\vspace{.13in}\noindent{\bf Remark 1:}
The key new element in this result is the fact that one can use,
for any ordering on the ground set $[n]$, the conditional values of
$f$ that appear in the upper and lower bounds for $f([n])$. Because
of \eqref{condred}, this is an improvement over simply using $f$.
The latter weaker inequality has been implicit in the cooperative
game theory literature; 
various historical remarks explicating these connections are given in
Section~\ref{sec:hist}. 
\par\vspace{.2in}

\vspace{.13in}\noindent{\bf Corollary I:} 
Let $f:2^{[n]}\ra\RL$ be any submodular function with $f(\nullset)=0$,
such that $f([j])$ is non-decreasing in $j$ for $j\in [n]$.
Then, for any collection $\collS$ of subsets of $[n]$,
\ben
\sumS \bs f (\setS | \setS^{c}\setminus >\setS) 
\leq f([n]) \leq  
\sumS \as f (\setS | <\setS)\, ,
\een
where $\beta$ is any fractional packing  and $\alpha$ is any fractional
covering of $\collS$.
\par\vspace{.13in}

\begin{proof}
The proof is almost exactly the same as that of Theorem I; the only difference being
that the validity there of (d) for fractional coverings and of (e) for fractional
packings is guaranteed by the non-negativity of $f(j|<j)$.
\end{proof}
\par\vspace{.13in}

Observe that if $f$ defines a polymatroid (i.e., $f$ is not only submodular
but also non-decreasing in the sense that $f(\setS)\leq f(\setT)$ if
$\setS\subset\setT$), then the condition of Corollary I is automatically satisfied.

\section{Entropy Inequalities}
\label{sec:main}

\subsection{Strong Fractional Form}

The main entropy inequality introduced in this work is the following
generalization of Shannon's chain rule. 

\vspace{.13in}\noindent{\bf Theorem I':}[{\sc Strong fractional form}]
For any collection $\collS$ of subsets of $[n]$, 
\ben
\sumS \bs H (\Xs | X_{\setS^{c}\setminus >\setS}) 
\leq H(X_{[n]}) \leq  \sumS
\as{H (\Xs | X_{<\setS})}\,
\een
and
\ben
\sumS \gs h (\Xs | X_{\setS^{c}\setminus >\setS}) 
\leq h(X_{[n]}) \leq  \sumS
\gs{h (\Xs | X_{<\setS})}\, ,
\een
where $\beta$ is any fractional packing, $\alpha$ is any fractional
covering, and $\gamma$ is any fractional partition of $\collS$.
\par\vspace{.13in}

One can give an elementary proof of Theorem I' as a refinement of that given by
Llewellyn and Radhakrishnan for Shearer's lemma (see \cite{Rad03}).
However, instead of giving the proof in terms of entropy (which one may
find in the conference paper \cite{MT07:isit}), we have proved in Theorem I
a more general result that holds for the rather wide class of submodular set functions.
To see that Theorem I' follows from Theorem I, we need to check that
the joint entropy set function $f(\setS)=H(\Xs)$ is a submodular function
with $f(\nullset)=0$. The submodularity of $f$ is a well known
result that to our knowledge was first explicitly mentioned by Fujishige \cite{Fuj78},
although he appears to partially attribute the result to a 1960 paper
of Watanabe that we have been unable to find. It follows from the fact that
$H(\Xs)+H(\Xt)-H(X_{\setS\cup\setT})-H(X_{\setS\cap\setT})=I(X_{\setS\setminus\setT};X_{\setT\setminus\setS}|X_{\setS\cap\setT})$ 
is a conditional mutual information (see, e.g., Cover and Thomas \cite{CT91:book}),
which is guaranteed to be non-negative by Jensen's inequality.
To see that the ``correct'' definition of $f(\nullset)=0$,
note that the ``unconditional'' entropy $H(\Xs)$ should be
equal to $H(\Xs|X_{\nullset})$, but the latter is $H(\Xs)-H(X_{\nullset})$
by definition, which suggests that $H(X_{\nullset})=0$.

Again, we would like to stress the freedom given by Theorem I' in terms
of choice of ordering. For convenience of notation, we simply chose one 
labelling of the indices using the natural numbers and used the 
ordering $1<2<\ldots<n$, but one may equally well use another labelling
or ordering.

\vspace{.13in}\noindent{\bf Remark 2:}
It is natural to ask what choices of fractional packing and covering
optimize the lower and upper bounds respectively.
For a given collection of subset entropies, the optimal
choices are clearly the solution of a linear program.
Indeed, the best upper bound is obtained,
for $\ws=H(\Xs|X_{<\setS})$, by solving:
\begin{quote}
Minimize $\sumS\as\ws$ \\
subject to $\as\geq 0$ and $\sum_{\setS\in\collS,\setS\ni i} \as \geq 1$.
\end{quote}
When the subset entropies are all equal, this is just the
problem of optimal fractional covering discussed in Section~\ref{sec:hyp}.
\par\vspace{.2in}

\subsection{Strong Degree Form}

The choice of $\alpha$ as the degree covering and
$\beta$ as the degree packing 
in Theorem I' gives the strong degree form
of the inequality.

\vspace{.13in}\noindent{\bf Theorem II:}[{\sc Strong Degree Form}]
Let $\collS$ be any collection of subsets of $[n]$,
such that every index $i$ appears in at least one element of $\collS$.
Then
\ben\label{thm:sd}
\sumS  \frac{H (\Xs | X_{\setS^{c}\setminus>\setS})}{r_{+}(\setS)} 
\leq H(X_{[n]}) \leq  \sumS 
\frac{H (\Xs | X_{<\setS})}{r_{-}(\setS)} .
\een
If $\collS$ is quasiregular, then the above inequality also holds for $h$ in place
of $H$.
\vspace{.13in}
 
\noindent{\bf Remark 3:} This also proves Proposition I. Indeed,
since conditioning reduces entropy, Proposition I is just the
loose form of Theorem II obtained by dropping the conditioning
on $<\setS$ in the upper bound, and including conditioning
on $>\setS$ in the lower bound.
\vspace{.13in}

\noindent{\bf Remark 4:} The collections $\collS$ for which
the results in this paper hold need not consist of distinct sets. That is, one may
have multiple copies of a particular $\setS\subset [n]$ contained in
$\collS$, and as long as this is taken into account in counting the degrees
of the indices (or checking that a set of coefficients forms a fractional
packing or covering), the statements extend. 
We will make use of this feature
when developing applications to combinatorics in Section~\ref{sec:count}.
\vspace{.13in}

\noindent{\bf Remark 5:} 
Using the previous remark, one may
write down Theorem II with arbitrary 
numbers of repetitions of each set in $\collS$. 
This gives a version of Theorem I'
with rational coefficients, following which an approximation argument can be used to
obtain Theorem I'. This proof is similar to the one alluded to by
Friedgut \cite{Fri04} for the version without ordering. 
Thus Theorem II is actually equivalent to Theorem I'.
\par\vspace{.13in}

The strong degree form of the inequality generalizes
Shannon's chain rule. In order to see this, simply choose
the collection $\collS$ to be $\collS_{1}$, the collection
of all singletons. For this collection, Theorem II says
\ben
\Sumn H(X_{i}|X_{[n]\setminus \geq i}) \leq H(X_{[n]}) \leq \Sumn H(X_{i}|X_{<i}) ,
\een
which is precisely Shannon's chain rule  (see, e.g., Shannon \cite{Sha48} and 
Cover and Thomas \cite{CT91:book}), since the upper and
lower bounds are identical. Note in contrast the looseness
of the upper and lower bounds in \eqref{weak:s1},
which are tight if and only if the random variables
$X_{i}$ are independent.

Application of Theorem II to non-symmetric collections
is also of interest. For instance, choosing $\collS$ to be the class of all sets of $k$
consecutive integers yields $r_{-}=1$ and $r_{+}=k$. Thus
\be\label{cons}
\frac{H(X_{[n]})}{\sum_{j\in[n]} H(X_{[j:l(j)]}| X_{<j})} \in \bigg[\frac{1}{k} ,1\bigg] ,
\ee
where $l(j)=\min\{j+k-1,n\}$. These examples make it clear that Theorem II
is rather powerful and generalizes well known results in addition
to producing new ones.

\subsection{Weak Fractional Form}

Theorems I' and II can be weakened by removing the conditioning in the upper bound,
and adding conditioning in the lower bound; from the latter, one obtains
the weak degree form of Proposition I, and from the former, one obtains 
the weak fractional form of our main inequality.

\vspace{.13in}\noindent{\bf Proposition II:}[{\sc Weak Fractional Form}]
For any hypergraph $\collS$ on $[n]$,
\be\label{frac:weak} 
\sumS  \bs {H (\Xs | \Xsc)} \leq H(X_{[n]}) \leq
\sumS  \as {H (\Xs)} 
\ee
and
\be\label{frac:weak2} 
\sumS  \gs {h (\Xs | \Xsc)} \leq h(X_{[n]}) \leq
\sumS  \gs {h (\Xs)} ,
\ee
where $\beta$ is any fractional packing, $\alpha$ is any fractional
covering, and $\gamma$ is any fractional partition of $\collS$.
\par\vspace{.13in}

\noindent{\bf Remark 6:}
While the main inequality as stated in both its degree form (Theorem II)
and its fractional form (Theorem I') seems novel,
the bounds have been known to various levels of generality, as
pointed out in the Introduction.
In the discrete mathematics community, particular forms of the upper bound
have been well known ever since the introduction of Shearer's lemma
by Chung, Graham, Frankl and Shearer \cite{CGFS86} (see also Radhakrishnan \cite{Rad03}
and Kahn \cite{Kah01b}). In the level of generality of Proposition II, 
the fractional form was demonstrated by
Friedgut \cite{Fri04} in terms of hypergraph projections. 
Friedgut's proof of the upper bound is perhaps not as transparent as the one we give.
In the information theory community, both the upper and lower bounds 
of Proposition II have been
known for the special case of the hypergraphs $\collS_{k}$ (consisting
of all sets of $k$ elements out of $n$), since the work of Han \cite{Han78}
and Fujishige \cite{Fuj78}. In this paper, we unify and extend all of these
results. 
\par\vspace{.13in}

\noindent{\bf Remark 7:}
In the case of independent random variables, the
joint entropy $H(X_{\setS})=H(X_{\setS}|\Xsc)=\sum_{i\in\setS} H(X_{i})$
is additive. Thus in that case, for any quasiregular
hypergraph $\collS$, Proposition I  holds with equality,
and this is just Lemma III with $a_{i}=H(X_{i})$. Similarly,
thanks to Lemma I, Proposition II holds with equality for 
independent random variables when $\alpha=\beta$ is a fractional partition. 
\par\vspace{.13in}

We believe that both the degree
formulations of Proposition I and Theorem II, and the fractional formulations of
Theorem I' and  Proposition II are useful ways to think about these inequalities,
and that they pave the way to the discovery
of new applications. We illustrate this by using the degree formulation
to count independent sets in graphs in Section~\ref{sec:count},
and by using the fractional formulation to obtain new determinantal
inequalities in Section~\ref{sec:det}.

\section{An Application to Counting}
\label{sec:count}

\subsection{Entropy and Counting}

It is necessary to recall some terminology from graph theory. For our purposes, a graph
$G=(V,E)$ consists of a finite vertex set $V$ and a collection $E$
of two-element subsets of $V$ called edges (allowing repetition,
i.e., self-loops). Thus $G$ is a special case of a hypergraph,
each hyperedge having cardinality 2. Two vertices are said to be
adjacent, if there is an edge containing both of them. An
independent set of $G$ is a subset $V_{I}$ of $V$ such that no two
vertices in $V_{I}$ are adjacent.

Given a graph $F=(V(F), E(F))$, the set $\Hom(G,F)$ of homomorphisms from $G$ to $F$ is
defined as
\ben\begin{split}
\Hom(G,F) = \{ x:&V \to V(F)  \ \mbox{ s.t. } \\
&\ uv\in E \Rightarrow x(u)x(v) \in E(F)\} .
\end{split}\een
Let $K_{a,b}$ denote the complete bipartite graph between parts of sizes
$a$ and $b$ respectively. 

Shearer's lemma, and more generally, entropy-based arguments,
have proved very useful in combinatorics. Shearer's lemma was
(implicitly) introduced by Chung, Graham, Frankl and Shearer \cite{CGFS86},
and Kahn \cite{Kah01b} stated an extension using the more familiar
entropy notation. Recent applications of Shearer's lemma
to difficult problems (where counting bounds are a key step in obtaining
the results) include F\"uredi \cite{Fur96}, Friedgut and Kahn \cite{FK98},
Kahn \cite{Kah01a, Kah01b}, Brightwell and Tetali \cite{BT03}, and Galvin and Tetali \cite{GT04}.
Radhakrishnan \cite{Rad03} provides a nice survey of entropy ideas used
for counting and various applications; see also the book by Alon and Spencer \cite{AS00:book}. 

The general strategy of entropy-based proofs in counting is as follows:
\begin{itemize}
\item To count the number of objects in a certain class $\mathcal{C}$ of objects,
consider a randomly drawn object $X$ from the class and note that its entropy
is $H(X)=\log |\mathcal{C}|$.
\item Represent $X$ using a collection of discrete random variables,
and apply a Shearer-type lemma to bound $H(X)$ using certain subset entropies
for a clever choice of hypergraph dictated by the problem.
\item Perform an estimation of the resulting bound, using Jensen's inequality
if necessary. 
\end{itemize}

Below, we follow this direction
of work and demonstrate a counting application of the new inequality.
In particular, we use Theorem I' to bound the number of
independent sets of an arbitrary graph, the number of proper graph
colorings with a fixed number of colors, and more generally the
number of graph homomorphisms.

\subsection{Counting graph homomorphisms}

Using Shearer's entropy inequality as a key ingredient, Kahn \cite{Kah06:pvt} 
recently showed a bound on the number of independent
sets of a regular graph $G$, building on his earlier result \cite{Kah01b} for
bipartite, regular graphs. Kahn's proof extends in a straightforward way, as observed by
D. Galvin \cite{Gal06:pvt}, to also provide
an upper bound on the number of homomorphisms from a $d$-regular
graph $G$ to arbitrary graph $F$. Theorem IV below
extends the observations of Kahn and Galvin to bound the number
of graph homomorphisms from an arbitrary graph $G$ to 
an arbitrary graph $F$.

\vspace{.13in}\noindent{\bf Theorem III:}[{\sc Graph Homomorphisms}]
For any $N$-vertex graph $G$ and any graph $F$, 
\be\label{eqn:hom_bound}
|\Hom(G,F)| \le  \prod_{v \in V}
|\Hom(K_{p(v),p(v)}, F)|^{\frac{1}{d(v)}} \, , 
\ee 
where $p(v)$ denotes the number of vertices preceding $v$ in any ordering
induced by decreasing degrees.
\par\vspace{.13in}

\begin{proof}
Let $X$ be chosen uniformly at random from $\Hom(G,F)$. The random homomorphism
$X$ can be represented by the values it assigns to each $i\in V$, i.e.,
$X=(X(1),X(2),\ldots,X(n))=(X_{1},X_{2},\ldots,X_{n})$,
where $X_{i}\in V_{F}$. By definition, $X_{i}$ and $X_{j}$
are connected in $F$ if $i$ and $j$ are connected in $G$. 
We aim to bound $H(X)$ from above.  

Let $\prec$ denote an ordering on vertices according to the 
decreasing order of their degrees (ties may be broken, for instance,
by using an underlying  lexicographic ordering of $V$). 
For each $i\in V$, let
\ben
P(i) = \{j \in V : \{i,j\}\in E \mbox { and } j\prec i\}\, ,
\een
and define $p(i) = |P(i)|$. Consider the collection $\collS$ to be the collection of
$P(i)$, and in addition, $p(i)$ copies of singleton sets $\{i\}$,
for each $i$.  Then observe that each $i$ is covered by $d(i)$
sets in $\collS$, i.e., that the degree of $i$ in the collection $\collS$
is $r(i)=d(i)$. Indeed,  each $i$ appears in $d(i) - p(i)$ sets of
the form, $P(j)$, corresponding to each $j$ such that $i\prec j$
and $\{i,j\} \in E$, and once in each of the $p(i)$ singleton sets
$\{i\}$.

By the upper bound in Theorem II applied to this collection $\collS$, 
we have
\begin{eqnarray*}
H(X) & \leq & \sum_{i \in V} \frac{1}{\min_{j \in P(i)}d(j)} H\Bigl(X_{P(i)} | X_{\prec P(i)} \Bigr) \\
& & + \sum_{i \in V} \frac{p(i)}{d(i)}  H(X_{i}| X_{\prec i})\\
& \leq & \sum_{i\in V} \Bigl( \frac{1}{d(i)} H(X_{P(i)})  + \frac{p(i)}{d(i)}  H(X_{i}| X_{P(i)}) \Bigr) ,
\end{eqnarray*}
by relaxing the conditioning and by the fact that the chosen ordering makes $j\in P(i)$ 
imply $d(j)\geq d(i)$.

Let $q_{i}$ denote the probability mass function of $X_{P(i)}$, 
which takes its values in $\calX_{i}=\{x_{P(i)}: x\in \Hom(G,F)\}$.
In other words, $q(x_{P(i)})$ is the probability that $X_{P(i)}=x_{P(i)}$,
under the uniform distribution on $X$. 
Finally, let $R(x_{P(i)})$ be the number of values that
$X_{i}$ can take given that $X_{P(i)} = x_{P(i)}$,
i.e., the support size of the conditional distribution of $X_{i}$ 
given $X_{P(i)} = x_{P(i)}$.
Note that this is also the number of possible extensions of the partial
homomorphism on $P(i)$ to a partial homomorphism on $P(i) \cup \{i\}$.

Then
\ben\begin{split}
&H(X_{P(i)})  + p(i) H(X_{i}| X_{P(i)}) \\
&\leq  \sum_{x_{P(i)}\in \calX_{i}} \Bigl(q(x_{P(i)}) \log \frac{1}{q(x_{P(i)})} \\
&\quad + p(i) q(x_{P(i)})  H(X_{i} | X_{P(i)} = x_{P(i)})\Bigr)\\
&\leq \sum_{x_{P(i)}\in \calX_{i}}  
q(x_{P(i)}) \log \frac{R(x_{P(i)}))^{p(i)}}{q(x_{P(i)})} \\
&\leq \log \sum_{x_{P(i)}\in \calX_{i}}  R(x_{P(i)}))^{p(i)} ,
\end{split}\een
where $R(x_{P(i)})$ is the cardinality of
the range of $X_{i}$ given that $X_{P(i)} = x_{P(i)}$,
and we have bounded $H(X_{i} | X_{P(i)} = x_{P(i)})$ by $\log R(x_{P(i)})$, 
and the last inequality follows by Jensen's inequality. 
Thus 
\ben
H(X) \leq \sum_{i\in V} \frac{1}{d(i)} \log \Bigl(\sum_{x_{P(i)}\in \calX_{i}} R_{i} (x_{P(i)})^{p(i)}\Bigr) .
\een

The proof is completed by observing that, for any $i\in V$,
\be\label{bip-hom}
\sum_{x_{P(i)}\in \calX_{i}} R_i (x_{P(i)})^{p(i)}
\leq  |\Hom(K_{p(i), p(i)}, F)| \,.
\ee
Indeed, first note that every (partial) homomorphism $x_{P(i)}$ of  $P(i)$ for any graph $G$ 
(regardless of the ordering $\prec$) is trivially  a valid (partial) 
homomorphism of one side of $K_{p(i),p(i)}$, since each side of this 
bipartite graph has no edges and $|P(i)|=p(i)$.
Furthermore, for a valid $x_{P(i)}$, the number of extensions $R_i(x_{P(i)})$  
to $i$ is the same whether the graph is $G$ or $K_{p(i),p(i)}$, since it 
only depends on $F$. This proves \eqref{bip-hom}.
Note that the inequality \eqref{bip-hom}
can be strict, since there can be partial homomorphisms of one side 
of $K_{p(i),p(i)}$ to a given $F$ which are not necessarily valid 
while considering (partial) homomorphisms
from $G$ to $F$, since the induced graph on $P(i)$, for a given $i$, might have
some edges.
(This corrects the claim in \cite{GT04} that \eqref{bip-hom} holds with equality.)
\end{proof}
\par\vspace{.13in}

Nayak, Tuncel and Rose \cite{NTR06} note that zero-error source-channel codes are
precisely graph homomorphisms from a ``source confusability graph''
$G_{U}$ to a ``channel characteristic graph'' $G_{X}$. Thus,
Theorem IV may also be interpreted as giving a bound on the number
of  zero-error source channel codes that exist for a given source-channel pair.

\subsection{Counting independent sets}

By choosing appropriate graphs $F$, various corollaries can be obtained.
In particular, it is well known that the problem of counting independent sets
in a graph can be cast in the language of graph homomorphisms.
Choose $F$ to be the graph on two vertices joined by an edge, and
with a self-loop on one of the vertices. 
Then, by considering the set of vertices of $G$ that are mapped
to the un-looped vertex in $F$, it is easy to see that each
homomorphism from $G$ to $F$ corresponds to an independent
set of $G$. This yields the following corollary.

\begin{figure}
\begin{center}
\includegraphics[angle=-89]{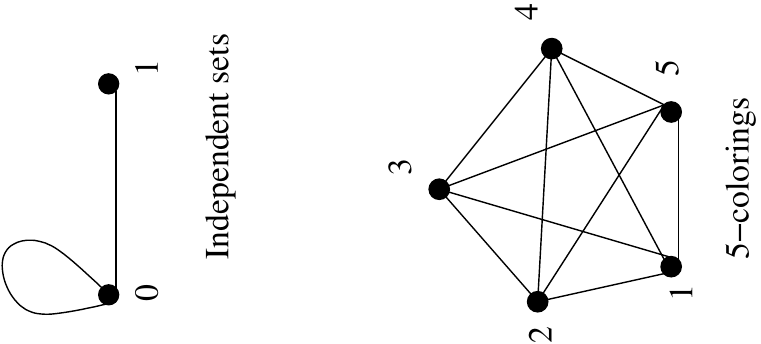}
\caption[]{The graphs $F$ relevant for counting independent sets
and number of $5$-colorings.}
\label{fig1}
\end{center}
\end{figure}

\vspace{.13in}\noindent{\bf Corollary II:}[{\sc Independent Sets}]
Let $G=(V,E)$ be an arbitrary graph on $N$
vertices, and let $\I(G)$ denote the set of independent sets of $G$. 
Let $\prec$ denote an ordering on $V$ according to
decreasing order of degrees of the vertices, breaking ties
arbitrarily.  Let $p(v)$ denote the number of neighbors
of $v$ which precede  $v$, under the $\prec$ ordering.  Then 
\ben
|\I(G)| \le \prod_{v \in V} 2^{(p(v)+1)\frac{1}{d(v)}} . 
\een
\par\vspace{.13in}

Specializing to the case of $d$-regular graphs  $G=(V,E)$  on $n$ vertices, 
it is clear that
\ben 
|\I(G)| \le \prod_{v \in V} 2^{(p_a(v)+1)\frac{1}{d}} \le
2^{\frac{n}{2} + \frac{n}{d}}. 
\een
where $\prec_a$ is an arbitrary total order  on $V$, and $p_a(v)$ is the
number of vertices preceding $v$ in this order, which are neighbors of
$v$. This recovers Kahn's unpublished result \cite{Kah06:pvt} 
for $d$-regular graphs, which generalized his earlier result \cite{Kah01b}
for the  $d$-regular, bipartite case. Note that we removed the assumption 
of regularity in Kahn's result by making a choice of ordering.


There is another way to view this result that is useful in
computational geometry. Namely, if one considers
a region (of, say, Euclidean space) and a finite family of subsets
$\calF=\{A_{v}:v\in V\}$
of this region, then one can define the intersection graph $G_{\calF}$
of this family by connecting $i$ and $j$ in $V$
if and only if $A_{i}\cap A_{j}\neq \nullset$.
Then the independent sets of $G_{\calF}$ are in
one-to-one correspondence with packings of the region
using sets in the family $\calF$. Thus Corollary II  also
gives a bound on the number of packings of a region using
a given family of sets.

Another easy corollary of Theorem III is to graph colorings. Recall that a
(proper) $r$-coloring of the vertices of $G$ is a mapping $f:V\to
[r]$ so that $u,v\in V$ and $uv \in E$ implies that $f(u) \neq
f(v)$. Consider the constraint graph be $F=K_r$, a complete graph
on $r$ vertices, for $r\ge 2$. Then $\Hom(G,K_r)$ corresponds to
the number of (proper) $r$-colorings of the vertices of $G$. Thus
the above theorem yields a corresponding upper bound on the number
of $r$-colorings of a graph $G$, by replacing 
$\Hom(K_{p(v),p(v)},F)$  in \eqref{eqn:hom_bound} with the number of
$r$-colorings of the complete bipartite graph $K_{p(v), p(v)}$.

\section{An Application to Determinantal Inequalities}
\label{sec:det}

The connection between determinants of positive definite
matrices and multivariate normal distributions is classical.
For example, Bellman's text \cite{Bel60:book} on matrix analysis
makes extensive use of an ``integral representation'' of determinants
in terms of an integrand of the form $e^{-<x,Ax>}$, which is 
essentially the Gaussian density. The classical determinantal inequalities 
of Hadamard and Fischer then follow from the subadditivity of entropy.
This approach seems to have been first cast in probabilistic language by 
Dembo, Cover and Thomas \cite{DCT91}, who further showed that
an inequality of Szasz can be derived (and generalized) using 
Han's inequality. Following this well-trodden path,
Proposition II yields the following general 
determinantal inequality.

\vspace{.13in}\noindent{\bf Corollary III:}[{\sc Determinantal Inequalities}]
Let $K$ be a positive definite $n \times n$ matrix and let $\collS$ be a
hypergraph on $[n]$. Let $K(\setS)$ denote the
submatrix corresponding to the rows and columns indexed by elements of $\setS$.
Then, using $|M|$ denote the determinant of $M$, 
we have for any fractional partition $\alpha^{*}$,
\ben
\prod_{\setS\in \collS} \Bigl(\frac{|K|}{|K(\setS^c)|}\Bigr)^{\alpha^{*}(\setS)}
\leq |K|
\leq \prod_{\setS \in \collS} |K(\setS)|^{\alpha^{*}(\setS)} .
\een
\par\vspace{.13in}

The proof follows from Proposition II via the fact that any positive definite 
$n \times n$ matrix $K$ can be realized as the 
covariance matrix of a multivariate normal distribution $N(0,K)$,
whose entropy is 
\ben
H(X_{[n]})=\half\log \big[(2\pi e)^{n} |K| \big] ,
\een
and furthermore, that if $X_{[n]}\sim N(0,K)$, then $\Xs\sim N(0,K(\setS))$.
Note that an alternative approach to proving Corollary III  would be to 
directly apply Theorem I to the known fact
(called the Koteljanskii or sometimes the Hadamard-Fischer inequality) that the
set function $f(\setS)=\log |K(\setS)|$ is submodular.

For an $r$-regular hypergraph $\collS$, using the degree partition 
in Corollary III implies that
\ben
|K|^{r} \leq \prod_{\setS \in \collS} |K(\setS)| .
\een
Considering the hypergraphs $\collS_{1}$ and $\collS_{n-1}$
then yields the Hadamard and prototypical Szasz inequality,
while the Fischer inequality follows by considering
$\collS=\{\setS,\setS^{c}\}$, for an arbitrary $\setS\subset [n]$.

We remark that one can interpret Corollary III using the all-minors 
matrix-tree theorem (see, e.g., Chaiken \cite{Cha82} or Lewin \cite{Lew82}).
This is a generalization of the matrix tree theorem of Kirchhoff \cite{Kir1847},
which states that the determinant of any cofactor of the 
Laplacian matrix of a graph is the total number of distinct spanning
trees in the graph, and interprets all minors of this matrix
in terms of combinatorial properties of the graph.

\section{Duality and Monotonicity of Gaps}
\label{sec:dual}

Consider the weak fractional form of Theorem I, namely 
\ben
\sumS  \gs f(\setS | \setS^c) \leq
f({[n]}) \leq \sumS  \gs f(\setS) . 
\een
We observe that there is a duality between the upper and lower bounds, 
relating the gaps in this inequality.

\vspace{.13in}\noindent{\bf Theorem IV:}[{\sc Duality of Gaps}]
Let $f:2^{[n]}\ra\RL$ be a submodular function with $f(\nullset)=0$. 
Let $\gamma$ be an arbitrary fractional partition using some hypergraph $\collS$ on $[n]$. 
Define the lower and upper gaps by
\be\begin{split}
\gap_{L}(f,\collS, \gamma) &=
f({[n]}) - \sumS  \gs f(\setS | \setS^c) \\
\mbox{ \ \ \ and \ \ \  }  
\gap_{U}(f,\collS, \gamma) &=
\sumS  \gs f(\setS) -  f({[n]}). 
\end{split}\ee
Then
\be\label{dual1}
\frac{\gap_{U}(f,\collS, \gamma)}{w(\gamma)} 
= \frac{\gap_{L}(f,\bar{\collS}, \bar{\gamma})}{w(\bar{\gamma})}\,,
\ee
where $w$ is the weight function and $\bar{\gamma}$ is the dual fractional partition 
defined in Section~\ref{sec:hyp}.
\par\vspace{.13in}

\begin{proof}
This follows easily from the definitions. Indeed,
\ben\begin{split}
&f({[n]}) - \sum_{\setS^c\in\bar{\collS}}  \bar{\gamma}(\setS^c) f(\setS^c | \setS) \\
=& f({[n]}) - \sumS  \frac{\gs}{w(\gamma) -1} \big[ f([n]) - f(\setS)\big] \\
=& \frac{\sumS\gs f(\setS)}{w(\gamma) -1} - \bigg[\frac{w(\gamma)}{w(\gamma) -1} -1\bigg] f([n]) \\
=& \frac{1}{w(\gamma) -1} \bigg[ \sumS\gs f(\setS) - f([n]) \bigg],
\end{split}\een
and
\ben\begin{split}
w(\bar{\gamma}) = \sum_{\setS^c\in\bar{\collS}} \bar{\gamma}(\setS^c) 
\,=\, \frac{\sumS \gs}{w(\gamma) -1}  
\,=\, \frac{w(\gamma)}{w(\gamma) -1}  .
\end{split}\een
Dividing the first expression by the second yields the result.
\end{proof}
\par\vspace{.13in}

Note that the upper bound for $f([n])$ with respect to 
$(\collS,\gamma)$  is equivalent to the lower
bound for $f([n])$ with respect to the dual $(\bar{\collS},\bar{\gamma})$, 
implying  that the {\it collection} of upper bounds for all hypergraphs and all fractional coverings
is equivalent to the {\it collection} of lower bounds for all hypergraphs
and all fractional packings. Also, it is clear that under the assumptions of
Corollary I, one can state a duality result extending Theorem IV by 
replacing $\gamma$ by any fractional covering $\alpha$, and 
$\bar{\gamma}$ by the dual fractional packing $\bar{\alpha}$. 

From Theorem IV, it is clear by symmetry that also
\be\label{dual2}
\frac{\gap_{L}(f,\collS, \gamma)}{w(\gamma)} 
= \frac{\gap_{U}(f,\bar{\collS}, \bar{\gamma})}{w(\bar{\gamma})} .
\ee
However, the identities \eqref{dual1} and \eqref{dual2}
do not imply any relation between $\gap_{U}(f,\collS, \gamma)$
and $\gap_{L}(f,\collS, \gamma)$.


The gaps in the inequalities have especially nice structure when they are 
considered in the weak degree form, i.e., for the fractional partition
using a $r$-regular hypergraph $\collS$, all of whose coefficients are $1/r$. 
The associated gaps are
\be\label{def:spl-gap}\begin{split}
g_{L}(f,\collS) &= f([n]) - \frac{1}{r} \sumS f(\setS | \setS^c) \\
\text{ and }\quad  
g_{U}(f,\collS) &= \frac{1}{r} \sumS f(\setS) -  f([n]) .
\end{split}\ee

\vspace{.13in}\noindent{\bf Corollary IV:}[{\sc Duality for Regular Collections}]
Let $f:2^{[n]}\ra\RL$ be a submodular function with $f(\nullset)=0$. 
For a $r$-regular collection $\collS$,
\ben
\frac{g_{L}(f,\bar{\collS})}{g_{U}(f,\collS)} = \frac{r}{|\collS|-r} .
\een
\vspace{.13in}


Let us now specialize to the entropy set function $\e(\setS)$-- we use this to mean 
either $H(\Xs)$ (if the random variables $X_i$ are discrete) or
$h(\Xs)$   (if the random variables $X_i$ are continuous).
The special hypergraphs $\collS_{k}$, $k=1,2,\ldots,n$,  consisting of all
$k$-sets or sets of size $k$, are of particular interest, and a lot is already known
about the gaps for these collections. For instance, 
Han's inequality \cite{Han78} already implies Proposition I for these hypergraphs, 
and Corollary IV applied to these hypergraphs implies that
\ben
\frac{g_{L}(\e,\collS_{n-k})}{g_{U}(\e,\collS_{k})} = \frac{k}{n-k} ,
\een
recovering an observation made by Fujishige \cite{Fuj78}. Indeed,
Theorem IV and Corollary IV generalize what \cite{Fuj78} interpreted
using the duality of polymatroids, since our assumptions are weaker
and the assertions broader. 
Fujishige \cite{Fuj78} considered these gaps important enough to
merit a name: building on terminology of Han \cite{Han78}, he called
the quantity $g_{U}(\e,\collS_{k})$ a ``total correlation'',
and $g_{L}(\e,\collS_{k})$ a ``dual total correlation''.

In two particular cases, the gaps have simple expressions
as relative entropies (see Section~\ref{sec:D} for definitions). 
First, note that the lower gap in Han's inequality \eqref{han:proto} is related to
the dependence measure that generalizes the mutual information.
\be\begin{split}
(n-1) g_{L}(\e,\collS_{n-1})
&= g_{U}(\e,\collS_{1}) \\
&= \sum_{i\in[n]} \e(\{i\}) - \e([n]) \\
&=D(P_{X_{[n]}}\|P_{X_{1}}\times\ldots\times P_{X_{n}} ) .
\end{split}\ee
It is trivial to see that the gap is zero 
if and only if the random variables are independent.

Second, the lower gap in Proposition I
with respect to the singleton class $\collS_{1}$ is  related to the upper gap in the
prototypical form \eqref{han:proto} of Han's inequality.
\be\label{eras-ent-gap}\begin{split}
g_{L}(\e,\collS_{1})&= (n-1)g_{U}(\e,\collS_{n-1}) \\
&= \sum_{i\in[n]} D(P_{X_{i}|X_{[n]\setminus i}} \| P_{X_{i}|X_{<i}} | P) .
\end{split}\ee
(Here the last equality comes from simple manipulation
of the pointwise log likelihoods.) Note that for the gap to be zero,
each of the relative entropies on the right must be zero. In particular,
$D(P_{X_{1}|X_{[2:n]}}\|P_{X_{1}})=0$, which implies
that $X_{1}$ is independent of the remaining random variables.
By applying the same fact to the collection of random variables
under different orderings, one sees that $X_{[n]}$ must be
an independent collection of random variables.

The latter observation is relevant to the study of the {\it erasure entropy}
of a collection of random variables, 
defined by Verd\'u and Weissman \cite{VW06:isit} to be
\ben
H^{-}(X_{[n]})= \Sumn H(X_{i}|X_{[n]\setminus i}) .
\een
They give several motivations for defining these quantities; most significantly,
the erasure entropy has an operational significance as
the number of bits required to reconstruct a symbol erased by an erasure channel.
Theorem 1 in \cite{VW06:isit} states that $H^{-}(X_{[n]})\leq H(X_{[n]})$
with equality if and only if the $X_{i}$ are independent. 
The inequality here is simply the lower bound of Proposition I
applied to the singleton class $\collS_{1}$, and is thus a special case of our results. 
The difference between the joint entropy of $X_{[n]}$ and its erasure entropy 
is just $g_{L}(\e,\collS_{1})$, and 
the characterization of equality in terms of independence follows from the 
remarks above. 
It would be interesting to see if the more general bounds on joint entropy
developed here can also be given an operational meaning using 
appropriate erasure-type channels.

Apart from the eponymous duality between the total and dual total correlations
discussed above, these quantities also satisfy a monotonicity property,
sometimes called Han's theorem (cf., \cite{Han78}). Since this complements the duality
result, we state it below in the more general submodular function setting.

\vspace{.13in}\noindent{\bf Corollary V:}[{\sc Monotonicity of Gaps}]
Let $f:2^{[n]}\ra\RL$ be a submodular function with $f(\nullset)=0$,
and let $g_{L}(f,\collS_{k})$ and $g_{U}(f,\collS_{k})$ be defined by \eqref{def:spl-gap}.
Then both $g_{L}(f,\collS_{k})$ and $g_{U}(f,\collS_{k})$
are monotonically decreasing in $k$.
\vspace{.13in}

\begin{proof}
Proposition I, applied to the collection $\collS_{k}$, immediately 
implies that $0=g_{U}(f,\collS_{n}) \leq g_{U}(f,\collS_{k})$,
for $k\in [n]$, on observing that
$r_{-}(\setS)=r_{+}(\setS)= \binom{n-1}{k-1}$.
To obtain the full chain of inequalities, first note that
for any $\setS$ in $\collS_{k+1}$,
\ben
f(\setS) \leq \frac{1}{k} \sum_{i\in\setS} f(\setS\setminus i). 
\een
Thus
\ben\begin{split}
&g_{U}(f,\collS_{k})-g_{U}(f,\collS_{k+1}) \\
&= \frac{1}{\binom{n-1}{k-1}} \sum_{\setS\in\collS_{k}} f(\setS) - 
\frac{1}{\binom{n-1}{k}} \sum_{\setS\in\collS_{k+1}} f(\setS) \\
&\geq \frac{1}{\binom{n-1}{k-1}}
\bigg[\sum_{\setS\in\collS_{k}} f(\setS) - \frac{1}{n-k} 
\sum_{\setS\in\collS_{k+1}} \sum_{i\in\setS} f(\setS\setminus i) \bigg] .
\end{split}\een
To complete the proof, note that
\ben\begin{split}
\sum_{\setS\in\collS_{k+1}} \sum_{i\in\setS} f(\setS\setminus i) 
&= \sum_{i\in [n]} \sum_{\setS\in\collS_{k+1}, \setS\ni i} f(\setS\setminus i) \\
= \sum_{i\in [n]} \sum_{\setS\in\collS_{k}, i\notin\setS} f(\setS)
&=\sum_{\setS\in\collS_{k}} \sum_{i\notin\setS} f(\setS) \\
=(n-k) \sum_{\setS\in\collS_{k}} f(\setS) &.
\end{split}\een
\end{proof}
\par\vspace{.13in}

Again specializing to the joint entropy function, let
\ben
\e_{k}^{(U)} = \frac{1}{\binom{n}{k}} \sum_{\setS:|\setS|=k} \frac{\e(\setS)}{k}
\een
denote the joint entropy per element for subsets of size $k$ averaged over
all $k$-element  subsets, and
\ben
\e_{k}^{(L)} = \frac{1}{\binom{n}{k}} \sum_{\setS:|\setS|=k} \frac{\e(\setS|\setS^c)}{k}
\een
denote the corresponding average of conditional entropy per element.
Since $g_{U}(\e,\collS_{k})=n\e_{k}^{(U)}-\e([n])$ and
$g_{L}(\e,\collS_{k})=\e([n])-n\e_{k}^{(L)}$,
Corollary V asserts that $\e_{k}^{(U)}$ is decreasing in $k$, while
$\e_{k}^{(L)}$ is increasing in $k$. 
Dembo, Cover and Thomas \cite{DCT91} give a nice interpretation of this fact,
briefly outlined below.
 
Suppose we have $n$ sensors collecting data relevant to the task at hand.
For instance, the sensors might be measuring the temperature of the ocean at
various points, or they might be evaluating the probability that a human face
is in a collection of camera images taken along the boundary of a high-security site,
or they might be taking measurements of neurons in a monkey's brain.
Suppose due to experimental conditions, at any time, we only have access to a random
subset of $m$ sensor measurements out of $n$. Then Han's monotonicity theorem
implies that, on average, we are getting more information as $m$ increases, etc.


\section{Entropy Power Inequalities}
\label{sec:epi}

Theorem I' implies similar inequalities for entropy powers.
Recall that the {\it entropy power} of the random vector $\Xs$ is
\ben
\calN(\Xs)=e^{\frac{2h(\Xs)}{|\setS|}} .
\een
This is sometimes standardized by a constant ($2\pi e$),
which is convenient in the continuous case as it allows for a comparison
with a multivariate normal distribution. For discrete random variables, one
can replace $h$ by $H$ in the above definition.

\vspace{.13in}\noindent{\bf Corollary VI:}
Let $\gamma$ be any fractional partition of $[n]$
using the hypergraph $\collS$. Then
\ben
\calN (X_{[n]})  
\leq  \sumS w_{\setS} \calN (\Xs) ,
\een
where $w_{\setS}=\frac{\gamma(\setS)|\setS|}{n}$ are weights
that sum to 1 over $\setS\in\collS$.
\par\vspace{.13in}

\begin{proof}
First note that
\ben
\sumS w_{\setS} =\sumS \frac{\gamma(\setS)}{n} \sum_{i\in\setS} \one_{i\in\setS}
= \sum_{i\in [n]} \nth \sum_{\setS\in\collS,\setS\ni i} \gamma(\setS) =1 ,
\een
since $\gamma$ is a fractional partition.
Thus
\ben\begin{split}
\exp\bigg\{\frac{2h(X_{[n]})}{n}\bigg\} 
&\leq \exp\bigg\{\frac{2}{n}\sumS \gamma(\setS)h(X_{\setS})\bigg\} \\
&=\exp\bigg\{\sumS w_{\setS} \frac{2h(X_{\setS})}{|\setS|}\bigg\} \\
&\leq \sumS w_{\setS} \calN (\Xs) ,
\end{split}\een
where the first inequality follows from Proposition II,
and the last inequality follows by Jensen's inequality.
\end{proof}

\vspace{.13in}\noindent{\bf Remark 8:}
Corollary VI generalizes an implication of Theorem 16.5.2 of 
Cover and Thomas \cite{CT91:book}, which looks at the
collections of $k$-sets. Note that, as in the special case covered in \cite{CT91:book}, 
Corollary VI continues to hold with the entropy power $\calN=\calN_{2}$ 
replaced throughout by any of the quantities
$\calN_{c}(\Xs)=\exp\{ch(\Xs)/|\setS|\}$
for any $c>0$. 
As in the case of entropy, the bounds on the entropy powers 
associated with the hypergraphs $\collS_{m}$ and the degree covering
satisfy a monotonicity property.
Indeed, by Theorem 16.5.2 of \cite{CT91:book},
\ben
\frac{1}{\binom{n}{m}} \sum_{\setS\in\collS_{n-m}} \calN_{c} (\Xs) 
\een
is a decreasing sequence in $m$.
\par\vspace{.13in}

More interesting than entropy power inequalities
for joint distributions, however, are entropy power inequalities
for sums of independent random variables with densities.
Introduced by Shannon \cite{Sha48}
and Stam \cite{Sta59} in seminal contributions, they have proved to
be extremely useful and surprisingly deep-- with connections
to functional analysis, central limit theorems, and to the determination
of capacity and rate regions for problems in information theory. Recently
the first author showed (building on work by Artstein, Ball, Barthe and 
Naor \cite{ABBN04:1} and Madiman and Barron \cite{MB07}) the following 
generalized entropy power inequality. For independent real-valued random variables
$X_{i}$ with densities and finite variances, 
\be\label{epi-std}
\calN\bigg(\sum_{i\in[n]} X_i\bigg)  \geq
\sumS \gs \calN\bigg(\sum_{i\in\setS} X_i\bigg) ,
\ee
for any fractional partition $\gamma$ with respect to any
hypergraph $\collS$ on $[n]$. Inequality \eqref{epi-std} shares an intriguing 
similarity of form to the inequalities of this paper, although it is much harder to prove. 

The formal similarity between results for joint entropy and for
entropy power of sums extends further. For instance, the fact that
\ben
\frac{1}{\binom{n}{m}} \sum_{\setS\in\collS_{n-m}} \calN\bigg(\sum_{i\in\setS} X_i\bigg)
\een
is an increasing sequence in $m$, can be thought of as a formal dual of
Han's theorem. It is an open question whether
upper bounds for entropy power of sums can be obtained that are
analogous to the lower bound in Theorem I'.

\section{An Inequality for Relative Entropy, and Interpretations}
\label{sec:D}

Let $A$ be either a countable set, or a Polish 
(i.e., complete separable metric) space equipped as usual with its Borel
$\sigma$-algebra of measurable sets. Let $\calP$ and $\calQ$ be
probability measures on the Polish product space $A^n$.
For any nonempty subset $\setS$ of $[n]$, write $\calP_{\setS}$ for the marginal
probability measure corresponding to the coordinates in $\setS$.
Recall the definition of the relative entropy:
\ben
D(\calP_{\setS}\|\calQ_{\setS})=E_{\calP} \bigg[\log\frac{d\calP_{\setS}}{d\calQ_{\setS}}\bigg] \in [0,\infty]
\een
when $\calP_{\setS}$ is absolutely continuous with 
respect to $\calQ_{\setS}$, and $D(\calP_{\setS}\|\calQ_{\setS})=+\infty$ otherwise.

One may also define the conditional relative entropy by
\be
D(\calP_{\setS|\setT}\|\calQ_{\setS|\setT} | \calP)=
E_{\calP_{\setT}} D(\calP_{\setS|\setT}\|\calQ_{\setS|\setT}) ,
\ee
where $\calP_{\setS|\setT}$ is understood to mean the conditional
distribution (under $\calP$) of the random variables corresponding
to $\setS$ given {particular values} of the random variables corresponding
to $\setT$; then $E_{\calP_{\setT}}$ denotes the averaging using $\calP_{\setT}$
over the values that are conditioned on.
With this definition, it is easy to verify the chain rule
\ben
d(\setS\cup\setT) = D(\calP_{\setS|\setT}\|\calQ_{\setS|\setT} | \calP) + d(\setT)
\een
for disjoint $\setS$ and $\setT$, so that following the terminology 
developed in Section~\ref{sec:submod}, we have
\ben
d(\setS|\setT)=D(\calP_{\setS|\setT}\|\calQ_{\setS|\setT} | \calP) .
\een
We have freely used (regular) conditional distributions in these
definitions; the existence of these is justified by the fact that
we are working with Polish spaces.

\vspace{.13in}\noindent{\bf Theorem V:}
Let $\calQ$ be a product probability measure on $A^n$, where
$A$ is a Polish space as above. Suppose $\calP$ is a probability measure on $A^n$
such that the set function $d:2^{[n]}\ra[0,\infty]$ given by
\ben
d(\setS)=D(\calP_{\setS}\|\calQ_{\setS})
\een
does not take the value $+\infty$ for any $\setS\subset [n]$.
Then $d(\setS)$ is supermodular.
\par\vspace{.13in}
\begin{proof}
For any nonempty $\setS,\setT\subset [n]$, we have
\ben\begin{split}
&d(\setS\cup\setT)+d(\setS\cap\setT)-d(\setS)-d(\setT) \\
&= [d(\setS\cup\setT)-  d(\setT)] - [d(\setS)-d(\setS\cap\setT)] \\
&= d(\setS\cup\setT\setminus\setT \,\big|\, \setT) - d(\setS\setminus\setS\cap\setT \,\big|\, \setS\cap\setT) .
\end{split}\een
Since $\setS\cup\setT\setminus\setT= \setS\setminus\setS\cap\setT$,
it would suffice to prove for disjoint sets $\setS'$ and $\setT$ that
\be\label{need-cvx}
d(\setS'|\setT) \geq d(\setS'|\setT') 
\ee
for any $\setT'\subset\setT$.

However observe that, since $\calQ$ is a product probability measure,
\ben\begin{split}
d(\setS'|\setT) = E_{\calP_{\setT}} D(\calP_{\setS'|\setT}\|\calQ_{\setS'})
= E_{\calP_{\setT'}} E_{\calP_{\setT\setminus\setT'}} D(\calP_{\setS'|\setT}\|\calQ_{\setS'})
\end{split}\een
and 
\ben\begin{split}
d(\setS'|\setT') = E_{\calP_{\setT'}} D(\calP_{\setS'|\setT'}\|\calQ_{\setS'})
= E_{\calP_{\setT'}}  D(E_{\calP_{\setT\setminus\setT'}} \calP_{\setS'|\setT}\|\calQ_{\setS'}) ,
\end{split}\een
so that \eqref{need-cvx} is an immediate consequence of the convexity
of relative entropy (see, e.g., \cite{CT91:book}). 
\end{proof}
\par\vspace{.13in}

Based on the supermodularity proved in Theorem V,
Theorem I applied to $-d(\setS)$ immediately implies
the following corollary.

\vspace{.13in}\noindent{\bf Corollary VII:}
Under the assumptions of Theorem V,
\be\label{mainD}\begin{split}
\sumS \gs &D (\calP_{\setS | \setS^{c}\setminus >\setS}\| \calQ_{\setS} | \calP)
\geq D(\calP_{[n]}\|\calQ_{[n]}) \\
&\geq  \sumS \gs D (\calP_{\setS | <\setS}\| \calQ_{\setS} | \calP)  ,
\end{split}\ee
where $\gamma$ is any fractional partition using any 
hypergraph $\collS$ on $[n]$.
\par\vspace{.13in}

\noindent{\bf Remark 9:}
We mention a hypothesis testing interpretation for the 
following easier-to-parse corollary of Corollary VII: for 
$r$-regular hypergraphs $\collS$ on $[n]$,
\be
D(\calP_{[n]}\|\calQ_{[n]})
\geq \rth \sumS D (\calP_{\setS}\| \calQ_{\setS} ) .
\ee
Suppose $\calP$ and $\calQ$ are two competing hypotheses for the
joint distribution of $X_{[n]}$. Then it is a classical fact due to Chernoff
(see, e.g., Cover and Thomas \cite{CT91:book}, where it is called Stein's lemma) that the
best error exponent for a hypothesis test between $\calP$ and $\calQ$
based on a large number of i.i.d. observations of the random
vector $X_{[n]}$ is given by $D(\calP_{[n]}\|\calQ_{[n]})$.
One may ask the following question: If one has partial access to all
observations (for instance, one observes only $\Xs$ out of
each $X_{[n]}$), then how much is our capacity to distinguish between
the two hypotheses $\calP$ and $\calQ$ worsened?  Corollary VII
can be interpreted as giving us estimates that relate our capacity to distinguish between
the two hypotheses given all the data to our capacity to distinguish between
the two hypotheses given various subsets of the data.
\par\vspace{.13in}


Interestingly, Corollary VII implies a tensorization property of the
entropy functional $\Ent_{\calQ}(f)=E_{\calQ}[f\log f]-(E_{\calQ}f) \log (E_{\calQ}f)$,
defined for positive functions $f$.
From the special case of Corollary VII corresponding to Han's inequality
(i.e., the hypergraph $\collS_{n-1}$), 
one obtains the classical tensorization property, as noticed by
Massart \cite{Mas00}.
We present below a generalized tensorization 
inequality for the entropy functional with respect to a product measure
by utilizing the power of Corollary VII more fully.

\vspace{.13in}\noindent{\bf Corollary VIII:}
Let $\collS$ be an $r$-regular hypergraph on $[n]$.
Then
\ben
\Ent_{\calQ_{[n]}}(g) \leq 
\frac{1}{r} E_{\calQ}
\sumS  \Ent_{\calQ_{\setS}}(g)  
\een
\par\vspace{.13in}

We omit the proof, which is based on the observation that
$\Ent_{\calQ}(f)=(E_{\calQ}f)\, D(\calP\|\calQ)$,
where $\calP$ is the probability measure such that
$\frac{d\calP}{d\calQ}=\frac{f}{E_{\calQ}f}$, and follows
the same line of argument as in \cite{Mas00}.

The tensorization property of the entropy functional 
is of enormous utility in functional analysis,
and the study of isoperimetry, concentration of measure, and convergence
of Markov processes to stationarity. For instance, see 
Gross \cite{Gro75}, Bobkov and Ledoux \cite{BL98},
and Kontoyiannis and Madiman \cite{KM06}, where
the classical tensorization property is used to prove logarithmic Sobolev inequalities for Gaussian,
Poisson and compound Poisson distributions respectively.

\section{Historical Remarks}
\label{sec:hist}

It turns out that the main technical result of this paper,
Theorem I, is related to a wide body of work 
in a number of fields, including the study of combinatorial optimization of set functions
in computer science, the study of cooperative games in economics, the
study of capacities in probability theory, and of course the study of structural properties of entropy in 
information theory, which has been our present focus. 
In this section, we sketch these connections and place our work in context.

The following terminology is useful.

\vspace{.13in}\noindent{\bf Definition IX:}
The set function $f$ is {\em fractionally subadditive} if 
\be\label{eq:fsa}
f([n]) \leq \sumS \gs f(\setS)\,,
\ee
for any $\collS \subset 2^{[n]}$, and 
for any fractional partition $\gamma: \collS\ra\Rpl$ of $[n]$.
If the inequality is reversed, we say $f$ is fractionally superadditive.
\par\vspace{.13in}

Note that Theorem I has the following corollary (basically
Proposition II for general submodular functions), obtained
by using \eqref{condred} to weaken the upper bound in Theorem I.

\vspace{.13in}\noindent{\bf Corollary IX:}
If $f$ is submodular and $f(\nullset)=0$, then it is fractionally subadditive.
\par\vspace{.13in}

This result has a long history, and has rarely been explicitly stated
in the literature although aspects of it have been rediscovered
on multiple occasions in various fields. First we describe how it is implicit in the classical theory
of cooperative games.

In cooperative game theory, a set function $f : 2^{[n]} \to \Rpl$ is called a value function;
it can be thought of as describing the payoff that can be obtained by arbitrary
coalitions of $n$ players, and it is canonical to take $f(\nullset)=0$. 
Different assumptions on the value function
$f$ correspond to different kinds of games. 
For instance, a {\it balanced game} is one for which the value function is 
fractionally superadditive, i.e.,
\be\label{bal-pty}
f([n]) \geq \sumS \gs f(\setS)
\ee
holds for every fractional partition $\gamma$.
If the value function $f$ is supermodular, the corresponding
game is said to be a {\it convex game}.

One solution concept for cooperative games 
is the core, a subset of Euclidean space representing possible allocations
of the payoff to players. (We do not bother to define it here; it suffices for our brief remarks
here to know that such a notion exists.) The fundamental Bondareva-Shapley theorem \cite{Bon63,Sha67} states that
the game with transferable utility associated with the value function $f$
has a non-empty core if and only if it is balanced.
Separately, it is known from even earlier work of Kelley \cite{Kel59} 
(see also Shapley \cite{Sha71} who rediscovered it in the language of games) 
that a convex game has a non-empty core. Putting these together, one sees 
that a convex game must be balanced. This yields a statement very similar to that
of Corollary IX.


Much more recently, yet another related approach to the relationship between
submodularity and fractional subadditivity has come from the 
theory of combinatorial auctions.
Lehmann, Lehmann and Nisan \cite{LLN01} 
showed that every submodular function is ``XOS'' (terminology that 
again we do not bother to explain here).
Feige \cite{Fei06} showed that XOS and
fractionally subadditive are identical. We refer the reader
to the mentioned papers for definitions and details.

To summarize, the literature from cooperative game theory and combinatorial
auction theory imply Corollary IX.

While we had expected direct proofs of Corollary IX to exist in the literature,
we had initially been unable to find a reference. After the first version of 
this paper was written and presented at various venues, we were informed by Alan Sokal that it
has indeed been explicitly stated and proved in the French statistical physics
literature by Moulin Ollagnier and Pinchon \cite{MP82} (see also van Enter,
Fern\'andez and Sokal \cite{VFS93}, where it is applied to entropy in a statistical
physics context).


The above discussion is also related to the theory of polymatroids. 
A nondecreasing and submodular set function $f : 2^{[n]} \to \Rpl$ 
with $f(\nullset) = 0$ is sometimes called a $\beta$-function.
This class of functions has been intensely studied ever since the pioneering
work of Edmonds \cite{Edm70}, who used them to define
polymatroids. Note that the nondecreasing property (i.e.,
$f(\setT) \leq f(\setS)$ whenever $\setT\subset\setS\subset [n]$)
implies that $f$ is non-negative. It is pertinent to note that the extra properties
inherent in polymatroid theory are not required for Corollary IX and 
Theorem I (for instance, a non-negativity requirement for $f$ would
rule out an application to the differential entropy); so Theorem I is
really just a basic fact about submodular functions.

\section{Discussion}
\label{sec:disc}

The inequalities presented in this note are contributions to a large body
of work on the structural properties of the entropy function 
for joint distributions. While the origins of such work clearly lie in Shannon's
foundational paper, let us again mention (see also the discussion after Theorem I') 
that the important observation of submodularity of the joint entropy function goes back 
at least to Fujishige \cite{Fuj78}. There have also been interesting new developments
in the last few years, namely the discovery of the 
so-called ``non-Shannon inequalities''. Motivated by the goal of characterizing
the possible joint entropy set functions $\e(\setS)=H(\Xs)$ for the discrete entropy
as the underlying joint distribution is varied arbitrarily, 
Zhang and Yeung \cite{ZY98} revealed a fascinating phenomenon: if 
one thinks of each such $\e$ (corresponding to any joint distribution on $n$
copies of a discrete alphabet) as being a vector of dimension $2^{n}$, then
the set of vectors one obtains in this manner is a strict subset of the set of vectors
corresponding to polymatroidal functions for any $n\geq 4$. The constraints
on joint entropy that are not automatic consequences of a polymatroid property
were termed ``non-Shannon inequalities'' in \cite{ZY98}. For more recent developments on this subject, 
one may consult Ibinson, Linden and Winter \cite{ILW06:isit},
Mat\'u$\check{\text{s}}$ \cite{Mat07}, or Dougherty, Freiling and Zeger \cite{DFZ07}.

In the context of these works, it is pertinent to note that all of the inequalities in 
this paper are Shannon inequalities, in the sense that they follow from
submodularity of an entropy function. Indeed, our study was based on the 
set function $\e(\setS)=H(\Xs)$, from consideration
of which our main entropy inequality (Theorem I') was derived.
However, since we now know from the mentioned literature that entropy satisfies additional constraints
beyond submodularity, a natural question arises.
If it is true that the set function
$\bar{\e}(\setS)=H(\Xs|X_{<\setS})$ is itself submodular, so that
Theorem I' then follows by an application of Corollary IX to $\bar{\e}$
rather than an application of Theorem I to $\e$, then we would
have a tighter outer bound on the space of joint entropy set functions.
The following counterexample shows that this is not the case.

\vspace{.13in}\noindent{\bf Proposition III:}
The set function $\bar{\e}(\setS)$ 
is not submodular.
\par\vspace{.13in}

\begin{proof}
We construct a counterexample with $n=4$ random variables.
Consider the sets $\setS=\{1,3\}$
and $\setT=\{3,4\}$. Then
$\setS\cup\setT=\{1,3,4\}$ and
$\setS\cap\setT=\{3\}$.
If $\bar{\e}$ is submodular, then since $\setS$ contains the first element,
\ben
H(\Xs)+H(\Xt|X_{<\setT}) \geq H(X_{\setS\cup\setT}) + H(X_{\setS\cap\setT}|X_{<(\setS\cap\setT)}) ,
\een
which in our case becomes
\be\label{ex-sm}\begin{split}
H(X_{\{1,3\}}) &+ H(X_{\{3,4\}}|X_{\{1,2\}}) \\
&\geq H(X_{\{1,3,4\}}) + H(X_{\{3\}}|X_{\{1,2\}}) .
\end{split}\ee
By the chain rule, 
\ben
H(X_{\{1,3,4\}})=H(X_{\{1,3\}})+H(X_{4}|X_{\{1,3\}}) ,
\een
and 
\ben
H(X_{\{3,4\}}|X_{\{1,2\}}) =H(X_{4}|X_{\{1,2,3\}}) +H(X_{3}|X_{\{1,2\}}) ,
\een
so that \eqref{ex-sm} reduces to
\ben\begin{split}
H(&X_{\{1,3\}}) +H(X_{4}|X_{\{1,2,3\}}) +H(X_{3}|X_{\{1,2\}}) \\
&\geq H(X_{\{1,3\}}) +H(X_{4}|X_{\{1,3\}}) + H(X_{3}|X_{\{1,2\}}) ,
\end{split}\een
and thence simply to 
$H(X_{4}|X_{\{1,2,3\}}) \geq H(X_{4}|X_{\{1,3\}})$.
However, this is in general not true since conditioning reduces entropy,
and thus the hypothesis of submodularity is falsified.
\end{proof}
\par\vspace{.13in}


Note, however, that such a counterexample is only possible when $\setS\cup\setT$
is strictly smaller than the index set $[n]$.

The relationship between the inequalities for discrete and continuous entropy
in this paper is worth noting. Observe that a slightly more general class of inequalities
holds for discrete entropy as compared to differential entropy (for instance, only
fractional partitions are allowed in the differential entropy context in Theorem I');
however, this is not surprising and indeed follows from the equivalences
explored by Chan \cite{Cha03}. 

The structural properties of entropy discussed in this work are
not just of abstract interest. Some applications, to determinant inequalities
and counting problems, have already been mentioned in earlier sections.
The inequalities discussed also have close connections with
several classical multiuser information theoretic 
problems, including the Slepian-Wolf data compression problem
and the multiple access channel. In particular, for the Slepian-Wolf
problem where data from $n$ sources is to be losslessly compressed
in a distributed fashion, it is the set function $H(\Xs|\Xsc)$ rather than $H(\Xs)$
that plays the key role. Consequently, the {\it lower} bound in Theorem I'
has a crucial significance: it is equivalent to the existence of 
a rate point whose sum rate is the same as the rate achievable
for non-distributed compression (namely $H(X_{[n]})$), and is one way
of showing that no extra cost is paid in terms of asymptotic rate for the distributed
nature of the task. These connections merit a separate and more
detailed exploration, and are discussed along with
several other applications of cooperative game theory to 
multiuser problems in \cite{Mad08:game}. 

Chain rules for entropy and relative entropy have played an important role in information
theory since their recognition by Shannon. Here we have presented several
inequalities for information in joint distributions that go beyond the chain rules
but can also be thought of as deeper consequences of them. While these relate
the information in projections of a random vector onto different subspaces, 
more general inequalities can be formulated that apply to a rich class of functions
beyond projections (such as the sum), and these are described along with applications
to additive combinatorics and matrix analysis in the follow-up works \cite{MMT08:pre, Mad08:pre}.
We anticipate further extensions and applications of these inequalities in the future. 
.

%
%

\section*{Acknowledgment}
We are indebted to Andrew Barron for many useful discussions,  
and for the indirect influence of Andrew's joint work \cite{MB07} with
MM on entropy power inequalities. We thank
the organizers of the IEEE International Symposium on Information Theory 2006 in
Seattle where we met and initiated this work, and
Ravindra Bapat, 
Uriel Feige, Gil Kalai and Alan Sokal 
for help with references. PT is thankful to the Theory Group at
Microsoft Research for hosting him during the period this research
was carried out. We are also deeply indebted to three anonymous referees
for very thorough feedback that eliminated an error and significantly 
improved the paper.



%

\end{document}